\newcommand\cH{{\cal H}}
\documentclass{article}
\usepackage[hypertex]{hyperref}
\usepackage{amsfonts}
\newcommand\lC{{\mathbb C}}

\title{How Free Will Could Will}
\author{George Svetlichny\footnote{Departamento de Matem\'atica, Pontif\'{\i}cia Universidade Cat\'olica, Rio de Janeiro, Brazil \newline
svetlich@mat.puc-rio.br \hfill \url{http://www.mat.puc-rio.br/\~svetlich}}}
\begin{document}
\maketitle

\begin{abstract}Many have proposed that free will would use quantum indeterminism. Strict adherence to the Born rule, which follows from the no-signal condition, seems to block this possibility. I propose here that if state collapse really does occur then there is a further form of indeterminism occurring in multipartite systems in that the basis upon which the collapse is to occur could be ambiguous. The choice of this basis is not covered by quantum mechanics nor subject to probability constraints and this provides a ground for a physical and eventually a mathematical model of free will
\end{abstract}
\vskip 12pt
{\obeylines
\hfill{\it La donna \`e mobile.}
\hfill{\emph{Rigoletto} Giuseppe Verdi}

}

\section{Introduction}

The existence of free will has been debated for millennia. Current biological and cognitive science research has amassed impressive apparent evidence that free will is just an illusion  \cite{cash:PNAS107.4499} and there seems to be little if any concrete scientific arguments to the contrary. Appeal to quantum indeterminism offers an apparent way that free will can exist, but up to now this suggestion has not been based on any concrete physical model. It also suffers from a seemingly fatal flaw. Quantum indeterminism is held to follow strict obeyance to definite probability distributions, the so called Born rule. If nature is playing dice, then the illuded  subject, thinking he is making a free choice of actions, is simply following the outcome of the dice throw, again just a slave of physical \label{luck} circumstances.\footnote{In the philosophical literature this is know as the \emph{luck argument} \cite{kane:JofP96.217,Fran:PhilStud156.199}: whatever happened by chance, it happened by mere luck, and what luck brings about is certainly not under control and so could not have been willed. Thus \emph{compatibilists} not only argue that free will is compatible with determinism, but also that any indeterminism undermines free will.  \emph{Libertarians} on the other hand insist that indeterminism is \emph{necessary} for free will. That this debate has not reached any consensus just points out that one is free to chose either side. I adhere to the libertarian view in this essay.} A natural sounding proposal, made by many, is that the Born rule must be modified to allow some flexibility. However, Gillis  \cite{gill:FOP41.1757} has convincingly argued that the Born rule is a consequence of causality, in the form of the no-signal condition for quantum processes. This seems to finally eliminate any possibility of free will without introducing new radical physics at the elementary lever. Without such a modification and lacking an explicit model, physics seems to offer no new real insights that could unravel the rather entangled philosophical debate on the question. Here I argue that no truly radical physics is needed. A reasonable extension of the Copenhagen state reductions hypothesis applied to molecular systems in living organisms can provide an effective theory of free will, free of the usual philosophical meanders.\footnote{I'm sure some philosophers, who just might happen to read this, would disagree.}

The next section presents a speculative physical model of free will. It is based on the assumption that state collapse really does occur  and that there are ambiguous collapse scenarios in which the choice of the basis within which reduction occurs is a primitive act of free will. The third and final section ties this model in to a theory of consciousness, also based on quantum state reduction, which I had presented in a previous essay \cite{svet:arXiv:1104.2634}. The conclusion is that free acts must be unconscious and always precede our conscious awareness of the instant of their initiation or of our conscious decision to act. Nevertheless such acts are truly free and we should be held responsible for them.

\section{A Physical Model of Free Will}

To start off, I propose the existence of free willing agents, {\it Alice\/} and others, which if necessary can be named {\it Bob\/}, {\it Charlie\/}, {\it Diedre\/},  and so on. For concreteness' sake let's think of them as existing somewhere in the human body\footnote{I'm  singling out humans for narrative's sake. The same should apply to other organisms and to some forms of what we now consider inanimate matter.} and free to act on quantum states of molecular systems, also in the body. There may be various agents in the same body, so I'm not saying that {\it Alice\/} is a person Alice as she is known to her friends and enemies, but that  {\it Alice\/} is an agent that is a part of the phenomenal entity that is a person known to others as such. This is the reason I italicize  the names of the agents. Due to lack of proper scientific vocabulary, I'm anthropomorphizing the entities and will resort to metaphors to get the main points across.
I will also need to refer to a person whose name will be Zoe. She has {\it Alice\/}, {\it Bob\/} and the rest as parts of her phenomenal being. By ``free" I mean that {\it Alice\/} and {\it Bob\/}, and the rest, can take an action {\em sui generis\/} among a certain available set without any previous causal link to other entities. The buck starts with them. Part of this freedom is the idea that the sequence of actions taken by an agent is subject to no law, not even of probability, meaning that  no proposition which would distinguish {\it Alice\/}'s continuation of a sequence of her acts from any other sequence can be deemed to be true.\footnote{ There is more to the notion of freedom than this,  but it'll do for the present essay. Besides, as we shall see later, {\it Alice\/}, like a stinging bee, destroys herself in her act. So even to speak of a sequence would be only in relation to renewed copies of the agent.} I'm not claiming that the agents are conscious  of their actions. Zoe will say she is conscious of her decisions to act, but this runs against any series of experiments which show that neurological actions leading to motor activity, such as Zoe raising her arm, precedes, sometimes by seconds, her appraisal of the moment in which she made a conscious decision to move her arm  \cite{soon-etal:NatNeuro11.543}. This has led many to conclude that free will is an illusion, as the notion of a free will decision is generally taken to be ``a {\em conscious\/} and free decision to act thus". Some authors however do speak of {\em unconscious \/} free decisions  \cite{bode-etal:plosone6.e21612}. The whole question deserves a lot more scrutiny and I will return to it later, but at the moment a ``free will" will be an initial agent of unconstrained causality as described above.

There is a tendency in discussing these issues to embark, following one's undisciplined intuition,  on arguments that tend to lead to infinite regresses. This should be avoided at all costs. Thus one often says that an agent \emph{decides} to act and then acts. But the decision is another act, so to act, the agent must act to act, and then, either by the same token, must act to act to act and so forth, or else the decision to act is an act \emph{simpliciter}. But if we admit acts \emph{simpliciter} why not make the first one such? Thus I posit that the agents act \emph{simpliciter}. Zoe says she decides to act, but \emph{Alice} just acts. The conundrum is that Zoe's decision follows \emph{Alice}'s act. Given a certain theory of consciousness, I'll argue that it must always be so.

Free willing agents will be identified with certain quantum states of molecular system to be described below. They act by conditioning quantum state collapse, or quantum state \emph{reduction} as it is also called.
State reduction is usually associated with the measurement process, which is a common situation for free will to enter physical discourse. One says that the experimenter is truly free to choose which measurement he/she is to perform on the system to be observed. I shall argue that such freedom can be made plausible on physical grounds and in particular, molecular system can act this way.\footnote{We can already point to Charlie, the experimenter himself, as a molecular system that acts this way, but the molecular complexity of Charlie is such that one can easily claim either the contrary, or else that Charlie is not merely a molecular system undergoing ordinary physical behavior but something that transcends this. Besides, this presumed position begs the question.}

The main reason that free will is often associated with state collapse is that according to the traditional ``Copenhagen'' view, a ``measurement process''  leads to  a truly discontinuous and probabilistic change. Free will, as is popularly  conceived, is also a discontinuous change, an act without cause besides the mere \emph{will} to cause it. It is natural to try to found free will on state reduction.  There are serious difficulties with this idea as was pointed out in the introduction. To make any headway we must examine the measurement process with renewed attention.

The paradigmatic situation of collapse via measurement was describe by von Neumann  \cite{vonn:MFQM} and is generally the way it is described in the literature. It's useful to review this. One has two quantum systems, the measured one described by the state vector \(\phi\) and the measuring one described by \(\Psi_0\) where the subscript \(0\) indicates that it is in a ``ready state'' before any measuring is done. The measuring system possesses a set of privileged \emph{pointer position states}  \(\Psi_i\) which are orthonormal.\footnote{The \emph{identification} of these pointer states is still an ongoing debate.} These are in one-to-one correspondence to a set of orthonormal states \(\phi_i\) of the measured system. We start off with the product state \(\Lambda_0= \phi\otimes\Psi_0\) which undergoes the following unitary time evolution after a certain time interval has elapsed:
\begin{equation}\label{premeas}
 \Lambda_0 \mapsto \Lambda_t=\sum_ic_i\phi_i\otimes\Psi_i,
\end{equation}
for some complex coefficients \(c_i\).

Actually, the measuring instrument is  not truly a free system and is subject to interactions with the environment (``the rest of the universe'') something that in principle should also be taken into account. We do not analyze this further and just simply assume that the states \(\Psi_i\) are really states of the combined measuring system and the environment.

At a certain time instant, not specified by any theory, the joint state changes discontinuously, in a way again not specified by any theory, to \emph{one} of the terms in the sum, say \(\phi_i\otimes\Psi_i\), and this happens with probability \(\|c_i\|^2\), which is the Born rule. We shall call a state such as (\ref{premeas}) which is subject to collapse a {\em collapse module.}
Though the von Neumann paradigm is consistent with all existing empirical evidence, it itself has not been subject to detailed experimental verification. One can, nevertheless use it as a guide to formulate collapse scenarios not directly covered by the above description and I shall do so in formulating a theory and model of free will.

As far as we know Zoe \emph{could} be an automaton, and all her behavior \emph{could} be a simple result of physical laws, probability included, without any causal connection to her claimed phenomenal self with its doubts, decisions, and illusory free actions. All these \emph{could} be epiphenomena. Nothing in our present knowledge of physics can preclude such a situation. This means that in providing a physical model of free will we must \emph{posit} at which point it can operate. Such a step will fatally seem \emph{ad hoc} and one must be parsimonious. I followed some simple rules in  identifying this point. First, the agent's, such as \emph{Bob}'s, action has to somehow manage state reduction as generally this is the only place in physics where ontological indeterminism is introduced. Second, for such agents  I posit only one faculty; they are capable of only one thing and all agents do the same kind of thing. Also what they do cannot be done by other types of systems. Furthermore, they do not think,  have desires, or ponder their situation. They are mesoscopic molecular system and are phenomenally void. Third, I refrained from positing any hitherto unknown physical laws or processes. I see this as a sort of Copernican principle applied to our cognition. We are nothing special,  physically or phenomenally. Whatever physical underpinnings our cognition has, it must operate universally and be already present in simpler organisms and even in inanimate systems. If this physics were in any way radical,  it would already have been discovered among the countless thousands of physics experiments already performed on complex samples of matter. Fourth, though experiments show Zoe's free decisions to act follow \emph{Alice}'s acts, and so  Zoe's introspection about her free will is not entirely reliable, it must to some extent reflect the facts and can be used as a guide with a dash of salt. These rules will not logically lead to a proper identification, but help keep thoughts on a narrow track.

There are of course truly autonomous processes within Zoe's body that do not rely directly on phenomenal content and, even if probabilistic, can be considered as strictly following physical laws. I'll refer to all of these processes as the \emph{autonomous system.}

By assumption, a free acting agent such as \emph{Bob} deals with collapse modules. He cannot change the probabilities of collapse as this is governed by the Born rule. His action must be therefore be restricted to either create, destroy, or choose between collapse modules, or still, to that of \emph{preventing} a collapse to occur in one. Now he himself is a molecular system which may incorporate the collapse module he is in charge of or not. If not, he must have more than one faculty, some sort of ``knowledge" of the module and the ability to act. I've discarded this possibility in favor of the greatest possible simplicity. Thus the collapse module must be part of \emph{Bob} himself. Creation and destruction can be plausibly handled by the autonomous system, thus we should not take these acts as the characteristic ones of the agents. If  \emph{Bob} can somehow  prevent a collapse (of himself),  then states  such  as (\ref{premeas}) would be capable of determining by their own volition when a collapse happens. The time instant of a ``counter click" in a measurement is not explained by any theory at the moment, so this hypothesis cannot be immediately discarded, but it seems implausible to me. The fact that unstable particle decay with rather precise exponential distributions seems also to argue against this idea. For \emph{Bob} to act as a censor it seems we would need new physical laws, which I'm discarding. Finally, our introspective appraisal of our free actions is not that of  \emph{censoring} all but one of the possibilities that present themselves, but of actively making a \emph{choice} in favor of one. If \emph{Bob} makes a choice between two collapse modules then both should be a part of  him. As I now show this provides a viable solution.

In the von Neumann paradigm the measuring instrument is a macroscopic system and so \emph{a fortiori} is the environment being the rest of the universe. A mesoscopic system of \(N\) parts could possibly also be divided into a measured and  measuring system and parts that belong to the environment and this division may be ambiguous.
Specifically, if \(\cH_N\) is the mesoscopic system's Hilbert space then one could have
\begin{equation}\label{ambimes}
    \cH_N=\cH_m\otimes\cH_M\otimes\cH_E=\cH_{m'}\otimes\cH_{M'}\otimes\cH_{E'}
\end{equation}
where the index \(m\) corresponds to the measured system, \(M\) to the measuring system, that which has pointer position states, and \(E\)  to the environment, which should be adjoined to the ``rest of the universe''.
It is conceivable that within such a situation more than one expression of form (\ref{premeas}) holds simultaneously:
 \begin{equation}\label{bicol}
\Lambda_t= \sum_ic_i\phi_i\otimes\Psi_i=\sum_ic_i'\phi_i'\otimes\Psi_i'
 \end{equation}
with two different bases for the measured system and two sets of pointer states. We shall call such a state an \emph{ambiguous collapse module.}\footnote{There may even be more than two simultaneous expression of form (\ref{premeas}) but for the present argument two suffices.}
 This could even happen if the Hilbert spaced on both sides of (\ref{ambimes}) are the same. Strict equality in (\ref{bicol}) would probably make the set of all ambiguous collapse modules have measure zero, and so physically implausible, but I feel that near equality to some degree would work just as well for what I discuss shortly below. If the Hilbert spaces on both sides of (\ref{bicol}) are the same then such an equality is only possible if the partial trace (in relation to either Hilbert space) of the density matrix corresponding to \(\Lambda\) has a degenerate spectrum. This is obviously a measure zero condition.

The simplest, and paradigmatic, example of a state satisfying (\ref{bicol})\footnote{I'm not saying this state is an ambiguous collapse module. This depends on having pointer states. What makes a state a pointer state is its relation to the environment.}  is the notorious singlet state in \(\lC^2\otimes \lC^2\) of two qubits.\footnote{I use the same basis in each copy of \(\lC^2\).}
\begin{equation}\label{singlet}
    {\sqrt 2}\,\Omega=\phi_1\otimes\phi_1-\phi_2\otimes\phi_2 = \phi_1'\otimes\phi_1'-\phi_2'\otimes\phi_2'.
\end{equation}
The two partial traces of the corresponding density matrix are both \(\frac12 I\), obviously with degenerate spectrum.

 If collapse is to occur in the face of basis ambiguity, in relation to which basis should it? The situation expressed by (\ref{premeas}) does not guarantee a collapse, it is just a condition that predisposes it. In the face of ambiguity, the collapse
may happen one way or another. The important point here is that this choice need not be subject to a probability law, nor to any law at all. For a \emph{fixed} measuring device, the possible outcomes must be subject to the Born probability rule to be consistent with causality (the no-signal condition). The \emph{choice} of the measuring device however is not constrained by the no-signal condition. It is here that free choice is possible. \emph{Bob} freely chooses which of the two collapse systems acts. Note that by this I am not introducing any new physical principle or law, what I'm saying is that where physics provides no law, there \emph{is} no law, and it is the absence of any law of behavior that is the hallmark of free action.\footnote{It may be strange to posit a physical process that obeys no law, and even stranger that I, a physicist,  would do so, but this seems unavoidable in a physical theory of free will!} The free agent's acts are then \emph{choices simpliciter}. Quantum mechanical state collapse (through one basis or another) is then the consequence of this choice. This is no longer an act of the agents.

 \emph{Bob}'s choices must have behavioral consequences. This would be the case whenever the results of collapse in one basis have different biological consequences from collapse in another basis. A particular action potential may be triggered by one collapsed state \(\phi_i\otimes\Psi_i\) in one basis and not by any other collapsed state in either of the two basis, which would have different biological consequences, or none at all. This is how  \emph{Bob}'s choices can lead to different behavior.

The physical conditions necessary for the above scenario to hold transcend a bit the conventional understanding of what a ``measurement process" is.  To base free will on this model  requires that, besides collapse happening in interactions of a quantum systems with macroscopic measuring devices, it also must happen within mesoscopic systems under appropriate circumstances. This in principle is an empirical question, not yet established. It must also be true that the ambiguous division described above be possible for such systems. This is another empirical question but doesn't seem to be ruled out by anything we know.

Thus the free agent \emph{Alice} is now identified with an  ambiguous collapse module, nothing more, nothing less. She acts \emph{simpliciter} and is phenomenally void without plans, desires, ideas, or other cognitive functions. Such cognitive function belong to Zoe of which \emph{Alice} is a part. \emph{Alice} is simply an agent whose choices  are not governed by any rule. On the other hand she, and \emph{Bob, Charlie, Diedre} and the rest are also part of Zoe and they have to act somehow in concert to be the constituents of Zoe's free actions. Here one may feel a certain contradiction, how can \emph{Alice} and the rest be free and still their acts be correlated to Zoe's desires. Perhaps the best though not very accurate paraphrase at this point is to say that \emph{Alice is truly free but perfectly obedient}. In my freedom I can freely chose to be fully obedient to another agent. Entanglement provides a paradigm for this idea of obedience under freedom. In the singlet state (\ref{singlet}) there is no causal link between the two qubits so in a sense they are free, but if the first is projected onto \(\phi_i\) the second is also projected onto \(\phi_i\) in the other Hilbert space. The qubits collapse obediently on the corresponding states due to the entanglement of the singlet state.  This coordinated  behavior, due to quantum correlations, is what allows the italicized horde of free agent to act in unison, obedient to the correlations in the collective free agent. I shall designate this collective agent by \emph{ALICE}, italicized since it is not Zoe but a part of her, and all in capitals to distinguish her from \emph{Alice, Bob} and the rest which are parts of \emph{ALICE}. Technically we must think of \emph{ALICE}  as a multipartite quantum system with the parts sharing quantum resources such as entanglement or, what is more likely, quantum discord that make the whole assembly act in unison in a correlated way. Notice that I'm saying that  \emph{Alice, Bob} and the rest are obedient to \emph{ALICE} and not to Zoe. Also I do not posit any agency within \emph{ALICE} that commands her.  To posit such an agency,  a \emph{willer} if you please,  pushes the problem one level higher. I would then have to explain how that \emph{willer} is constituted and acts. In doing so, the mental habits that pushed me up to this level will kick in again and force me to go up yet another level of explanation and so ad infinitum and ad nauseam. One has to cut short this regress that one so easily falls into. Simply put, \emph{ALICE is} the free willing part of Zoe.

Note that by identifying \emph{Alice} with an ambiguous collapse module,  once she acts, she is gone, as the collapsed state is either no longer an ambiguous collapse module, or one that is not \emph{Alice}. For that particular action to happen again she must be resurrected, which could happen by the autonomous system, or possibly by some of her italicized colleagues.\footnote{A collapsed state of \emph{Bob could} be \emph{Alice}. Such type of recursion is not ruled out.} Thus  \emph{ALICE} is a highly dynamical entity with parts that come into and go out of existence all the time. \emph{ALICE} is fickle. If the individual agents were luminous and we could see them, \emph{ALICE} would be a shimmering swarm of lights. For such an entity to exist there must be constant activity in Zoe's body that sustains it.

\section{Free Will and a Quantum Collapse Theory of Consciousness}

Zoe  just finished writing the next chapter in her quantum crime mystery novel.  She wonders if she should now have lunch at home and continue working, or  should she visit Diedre, her  mathematician friend and science advisor, or maybe even just call it a day and spend the rest of it with Charlie, if he can be pried away from his laboratory. She decides to call  Diedre to clarify the idea of quantum money, something she  just put into her novel.

This is a prosaic moment in the life of a young novelist, but there are issues that have troubled thinking people for millennia. What was it exactly that set Zoe thinking about her possible actions? Was it a spontaneous uncaused happening, or was it a result of any number of previous physiological and mental processes occurring, possibly, over many days? Was it a muse? And the final decision to consult with Diedre, was this an act of free will as we may want to think, or was it also conditioned, if not caused by something else? A believer in free will will want to interject a free spontaneous act somewhere, but where? And if this act was conditioned by thoughts and feelings either at the time or earlier, can it still be consider free? We need a true physical model of free will and other phenomenal entities to begin answering all these questions and be able to do decisive experiments and mathematical analyses.

So to begin an analysis of how \emph{ALICE} is related to the rest of Zoe's  phenomenal content, especially her consciousness, one must have a model of this content. In principle any number of theories of consciousness can incorporate free will based on ambiguous collapse modules. I shall here only briefly consider a theory I had previously proposed in an essay entitled \emph{Qualia are Quantum Leaps} \cite{svet:arXiv:1104.2634} and that I will refer to as the \emph{Collapse Theory of Consciousness (CTC)}.\footnote{This acronym coincides with the one for \emph{Closed Time-like Curve} but I'm sure no confusion will ever arise. This coincidence is somehow satisfying.} The present essay can be considered as an extension of the previous one providing greater speculative detail concerning one aspect (free will) of phenomenal existence. In \emph{Qualia} I speculated that free will could act through modification of the Born rule. I now feel that that suggestion was misguided due to the relation of the Born rule to causality and many other aspect of quantum mechanics. If through an act of free will one could modify probabilities given by the Born rule then one could communicate superluminally. To avoid the luck argument  against indeterministic free will (see footnote on page \pageref{luck}) one has to find an  indeterminism  subject to neither probability nor the no-signal restraints.\footnote{I'm sure the results of such lawless acts would still be called by some as due to luck. Luck is a slippery concept, but is generally considered in a probabilistic setting. Calling any act without cause as being due to luck is to  stubbornly beg the question.} The ambiguous collapse modules satisfy this if they do exist. Their macroscopic versions seemingly do exist in our free choice to perform this or that experiment just on our whim. It is possible that scaled-down versions exits also, and this is what I have postulated lies at the heart of free will. According to CTC, Zoe's  phenomenal content (such as consciousness) \emph{is} the presence of quantum state reductions in her body (and possibly extension of it). Her free willed actions are the choices \emph{ALICE} makes upon the system of ambiguous collapse modules, which in reality is \emph{ALICE} herself.

So let us again, now from the CTC viewpoint, consider the moments of Zoe's life described above. There are state reductions going on constantly in and around her body. These constitute the phenomenal existence of Zoe, her feeling, thoughts, desires, all the so called \emph{qualia} in her consciousness and, as will be discussed shortly, all the unconscious phenomenal entities that make up Zoe as a subject.  One should not think that these processes are some sort of \emph{manifestation} of her self,  ``rendered'' in her body and that her self transcends and commands. It \emph{is} her self. There is nothing beyond or transcendent to these state reductions that is Zoe's self. I will call the full phenomenal and physical entity describe above by ZOE. \emph{ALICE} is obedient to ZOE but not to Zoe. Zoe \emph{thinks} \emph{ALICE} is obedient to her when she feels her decisions initiate free acts, but that is an illusion. Some of what in the previous section I've attributed to Zoe must, under CTC, be actually attributed to ZOE.

Because of the presence of classical and quantum correlations in such a system of state reductions, the whole system acts in concert. The thoughts, feelings, qualia, etc, at each moment modify both through quantum  and classical processes, the global quantum state, within which as a part is \emph{ALICE}, which is free to will. It is in this way that Zoe's thoughts, feelings, and past actions affects her present free acts by providing choices within particular ambiguous collapse modules.

Note that CTC is a form of physicalism. The whole realm of phenomenal existence is placed within the province of physics. Zoe in her phenomenal existence acts by physics. Her phenomenal existence is subject to precise mathematical description and analysis, still to be discovered. Her self is subject to the same empirical investigation as any other physical phenomena in this universe.

There is still the argument that neurological precursors of the conscious decision to act is an indication that free will is an illusion. This of course if based on the idea that free will need be a \emph{conscious} decision to act and anything not conscious is not free. The ground breaking experiment on unconscious precursors of conscious decisions was performed by Libet and collaborators \cite{libe-etal:Brain106.623}. In a subsequent publication \cite{libe:BBS8.529}, in an apparent attempt to save consciousness' role in free actions,  Libet suggested that consciousness would act as a \emph{censor} blocking the development of \emph{ALICE}'s act if Zoe  finds that it would have undesirable consequences. However there is no reason this censoring act by Zoe should not be preceded itself by a censoring act of \emph{ALICE} and we're back in the same boat. Otherwise we'd be piling one faculty of free will upon another and off again either on yet one more infinite regress or another qualitatively new mental capability. This begs for Occam's razor.

The present proposal does not require free will to be conscious.  Consciousness is just one aspect of the system of state reductions and free will is the system of ambiguous  collapse modules, \emph{ALICE}, a different aspect. Within the CTC perspective action must begin earlier than the conscious decision to act. Consciousness of a decision is the quantum collapse of a certain state, but \linebreak\emph{ALICE} acts \emph{prior} to any collapse. It is a \emph{choice simpliciter} of a basis for collapse made by a correlated swarm of phenomenally void agents.\footnote{Whether Zoe can be held morally responsible for \emph{ALICE}'s acts is a separate issue. In my opinion she can be, and no doubt ZOE can.} Zoe's conscious decision is akin to a \emph{report} of an action on \emph{ALICE}'s part.  Many neurobiological studies show that subjects react to external (and by extension must also react to internal) stimuli without being conscious of these. One has to distinguish somehow between \emph{unconscious awareness} and \emph{consciousness} strictly speaking, which is \emph{conscious awareness}. All processes in Zoe's body, insofar as they lead to specific state reductions, must be considered as part of phenomenal existence. All quantum state reduction in a highly correlated quantum state, of which \emph{ALICE} is part, is awareness, conscious or not. It is in this correlated existence that \emph{ALICE} obeys ZOE. Consciousness is only part of this awareness, almost an afterthought. Arguments have been brought forth that consciousness is a result of attention which is thought of as inscribing into working memory \cite{debri-prin:WIRCS1.51}. It may be that memory is much more complex than previously though and that consciousness is to some extent a form of memory, a ``remembering the present", as some have put it. There is currently gathering evidence toward such a view, see  \cite{meye:Science335.415} and references therein. Thus as I look out the window and see a green hillside, it's not that I'm conscious of a green hillside, it's that I have a memory of having been aware, unconsciously, of a green hillside.\footnote{More or less a few milliseconds prior.} This would be a very fast and very short lived memory, something like the RAM memory of a computer, which is constantly refreshed as I continue looking and  believing I'm conscious of green and a hillside. The massive amount of other information I'm aware of does not enter consciousness. Such a filter is a product of evolution for survival purposes. The whole collapse system is aware of a vast amount of information but only remembers a tiny fraction (including the content of consciousness) that is important for continued existence, the rest is discarded. Under such a concept, free will acts without consciousness, and when we think that we've made a decision and are conscious of this, it's just a memory of an earlier act. Such a free unconscious act has no time stamp and so we think it's origin is contemporary with our conscious awareness of it and not prior. This does not mean that our decisions (memories of acts) are not ours nor that they are not free. It just means that what our self is, is a much more complex physical system, of a phenomenal kind, than we had imagined before. Zoe knows herself through a narrative, one presented to her by ZOE who writes in Zoe's memory \emph{this is what you've seen and done,} and Zoe reads herself as a novel, one based on a true story, but a novel nevertheless.


\end{document}